\documentclass[11pt]{article}
\usepackage[paper=a4paper,left=30mm,right=30mm,top=25mm,bottom=25mm]{geometry}
\usepackage[utf8]{inputenc}
\usepackage{graphicx}
\usepackage{newpxtext,newpxmath} 
\usepackage{amsmath}
\usepackage{bbold} 
\usepackage{braket}

\usepackage[sorting=none,giveninits,url=false,date=year,isbn=false,doi=false]{biblatex} 
\usepackage{hyperref} 

\newcommand{\beq}{\begin{equation}}
\newcommand{\eeq}{\end{equation}}
\newcommand{\smallfrac}[2]{\begin{matrix} \frac{#1}{#2} \end{matrix}}
\newcommand*\sqs{\smallfrac{1}{\sqrt{2}}}

\title{Playing with a Quantum Computer}
\author{Rainer Müller$^1$ \& Franziska Greinert$^2$ \\ \small
Technische Universität Braunschweig\\ \small Bienroder Weg 82, D-38110 Braunschweig, Germany\\
\small $^1$rainer.mueller@tu-bs.de, $^2$f.greinert@tu-bs.de}

\date{\today}

\begin{document}
	
\maketitle

\section{Introduction}
The new quantum technologies are currently attracting a great deal of public attention. This application area of quantum physics is expected to provide significant technological and economic opportunities. Major projects such as the Quantum Flagship of the EU~\cite{QuantumFlagshipFuture} or the US National Quantum Initiative~\cite{NationalQuantumInitiative} have been launched with the aim of bringing quantum technologies into industrial application. The development of quantum computers, which is being vigorously pursued, is attracting particular interest. 
Journals and internet news channels regularly report on progress. 

The high public attention given to quantum computing shows that it is perceived as an interesting topic. 
We want to utilize this motivating effect for the teaching and learning of quantum physics. Specifically, we want to take advantage of the access to real quantum computers, which various providers make available free of charge.

A number of platforms (e.g. IBM Quantum~\cite{IBMQuantum} or TU Delft's Quantum Inspire~\cite{QuantumInspire}) allow users to register for cloud-based quantum computer access. In these environments, users can try out quantum algorithms on real quantum hardware. In addition, there are user-friendly simulators such as Quirk~\cite{Quirk, gidneyHowUseQuirk} and environments with extensive learning materials for learning hardware-related programming languages (e.g. IBM Qiskit~\cite{Qiskit}, Microsoft Q\#~\cite{QsharpENG} or Google Cirq~\cite{Cirq}). New providers and approaches appear regularly~\cite{matthewsHowGetStarted2021}. An overview of freely available quantum programming resources is provided by the regularly updated GitHub collection ``Open-Source Quantum Software Projects''~\cite{CuratedListOpensource2021}.

The approach of teaching quantum physics via quantum technologies has one major advantage: the basic entities are physically simple. Qubits are described as two-state systems -- the simplest possible quantum systems. The polarization of light, which plays a major role in quantum cryptography and communication, is also very easy to describe quantum mechanically. Another teaching advantage is that quantum technologies directly address the genuinely non-classical features of quantum physics: topics such as superposition, measurement, entanglement are essential in this area.

The physics underlying the new quantum technologies is not new: It is still the same
quantum physics developed in the days of Heisenberg and Schrödinger. But its technological application in quantum computers, with novel concepts as qubits and quantum gates, allows a new teaching approach to quantum physics. The new approach is focused more on information science than the traditional approach. It opens up new opportunities for 
application orientation and allows for fresh examples and exercise tasks. 

In this article we would like to show a direct and straightforward way to use quantum computers in an introductory course on quantum physics. The technical possibilities are already available: The platforms listed above provide learners with the opportunity to work with real quantum computers and program them in intuitively usable programming environments. 

There is an obstacle, however, to this new and motivating approach to quantum physics: 
Common quantum algorithms are complicated. Therefore, they often remain incomprehensible to learners. Quantum algorithms thus create a barrier to learning quantum computing. The simplest established quantum algorithm with a quantum advantage, the Deutsch-Josza algorithm \cite{deutschRapidSolutionProblems1992}, is easy to program. However, it has the disadvantage of solving a problem whose meaning is not immediately obvious to most people. Other prominent quantum algorithms, such as Shor's algorithm \cite{shorAlgorithmsQuantumComputation1994} and Grover's algorithm \cite{groverFastQuantumMechanical1996}, require difficult concepts such as modular arithmetic, the quantum Fourier transform, or the concept of amplitude amplification, making them difficult for learners. Of course, it is also possible to implement algorithms without quantum advantage. However, it seems unsatisfactory to use a quantum computer to implement, say, a half adder.

To overcome this difficulty, we use an algorithm that solves a simple and easily understandable problem while providing a quantum advantage. The algorithm we propose does not address a classical problem from computer science. Rather, it is a simple game in which the use of quantum physics offers a winning advantage. The game is called Quantum Penny Flip and was proposed by David A. Meyer back in 1999 \cite{meyerQuantumStrategies1999}. It can be easily reformulated to be described by quantum gates. We can therefore use it to teach the programming of a quantum computer

The Quantum Penny Flip game has already been implemented as a computer game, either individually \cite{zapirainQuantumPennyFlip2019, zapirainQuantumPennyFlip2020GitHub} or in a combination with other small quantum games \cite{alamLinkQuantum2019a}. In these implementations, however, the focus has been on playing and experiencing quantum phenomena, rather than programming a quantum computer.  
Furthermore, the computer games require software to be installed, which can be a hurdle for use in teaching. 

In this article, we describe how the Quantum Penny Flip game can be used for an introduction to gate-based quantum computing. The game can be used to easily and convincingly demonstrate a form of quantum supremacy. We use IBM's Quantum Composer~\cite{IBMQuantumComposer} for the implementation. The use of this environment does not require any installation. In addition, the graphical programming environment can be used without registration. A free account is required only if real quantum computer hardware is used.

Our target audience is students in introductory courses on quantum physics and also high school students who are interested in the topic. The mathematical prerequisite is an elementary understanding of Dirac's bra-ket formalism. Only the simplest basics of the formalism are needed, however. The goal is to take advantage of the motivation and learning opportunities provided by the access to a real quantum computer. In the following we will describe the Quantum Penny Flip algorithm and learn about some important quantum gates along the way.

\section{A simple coin toss algorithm}
Quantum Penny Flip is a game with two participants who take turns performing a sequence of moves. 
The starting situation is the classical coin flip with two players (Alice and Bob): a coin is tossed into the air. If it shows ``heads'' when it lands, Alice has won. If it shows ``tails'', Bob has won. To make the game adaptable to the quantum algorithm, we slightly reformulate the rules for the classical version of the game:

\begin{enumerate}
	\item Alice places the coin in a state of her choice (heads or tails on top) in a box that covers the coin. Even by touching the coin, it is impossible to determine which side is up. 
	\item Bob reaches into the box and has the choice to flip the coin over or not. Alice cannot see what he is doing.
	\item Alice reaches into the box and performs an operation of her choice (flip or not flip).
	\item The coin is uncovered and the result is read.
\end{enumerate}

\begin{figure}
	\centering
	\includegraphics[width=12truecm]{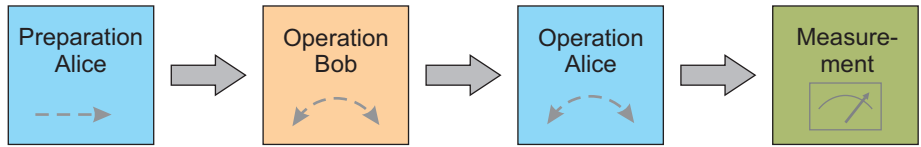}
	\caption{The Quantum Penny Flip game in the scheme preparation -- interaction -- measurement.}
	\label{abb5_7}
\end{figure}

\noindent
By formulating the rules of the classical version of the game in this way, we are already employing the scheme "preparation -- interaction (= operations) -- measurement" (Fig.~\ref{abb5_7}). This scheme is often used to describe quantum mechanical processes \cite{ballentineQuantumMechanicsModern1998}.

Even though the roles of Bob and Alice are not quite distributed symmetrically, it is clear that there is no winning strategy for either player. Chances are equal for both. Therefore, it is surprising that there is a clear winning strategy for Alice in the quantum version. In our approach, we want to utilize this amazing fact to motivate students to learn quantum physics. 

\section{Quantum version of the algorithm: Quantum Penny Flip}
The quantum version of the algorithm runs entirely according to the rules just established. The main difference is that the classical coin is replaced by a qubit. Qubits are the smallest basic unit of a quantum computer. They are described by two basis states $\ket{0}$ and $\ket{1}$. For the basic discussion of quantum algorithms, we do not need to know anything about the physical realization of these states. They may be polarization freedom lines of light, hyperfine states of trapped ions, or electron states in superconductors -- we can abstract from this for now. The requirements for the physical realization of qubits have been summarized by DiVincenzo in a set of five criteria \cite{divincenzoPhysicalImplementationQuantum2000}. In the present context, the most relevant of them is existence of long-lived superposition states. Our algorithm uses only a single qubit with two states. Thus, it is the simplest conceivable quantum system.

For our game, we assume that Bob is limited to the operations that are also possible in classical physics. The choice is small for him: he can flip or not flip the coin. In computer science, the flip of a bit is described by a NOT gate that swaps 0 and 1. In quantum computing, this is replaced by the Pauli $X$ gate, explained in more detail below. Not flipping the coin is described by the identity operation~$\mathbb{1}$.

Unlike Bob, Alice knows the laws of quantum physics and is able to perform all conceivable quantum operations on the qubit. To discuss this, we need to describe the qubit as well as the operations performed on it in the language of quantum physics. There are three ways to do this: (1) in the Dirac's bra-ket formalism, (2) by two-component vectors -- the gate operations are then described by $2 \times 2$ matrices -- and (3) geometrically using the Bloch sphere~\cite{rieffelQuantumComputingGentle2011}. All three possibilities are easy to implement; in this paper we take the first approach. 

The general state of the qubit is a superposition of the two orthogonal basis states:
\beq \ket{\psi} = \alpha \ket{0} + \beta \ket{1}. \label{eq1}\eeq
Here $\alpha$ and $\beta$ are complex coefficients with $|\alpha|^2 + |\beta|^2 = 1$. In quantum computing, it is generally assumed that each qubit is initially prepared in the state $\ket{0}$.

\section{Qubit operations}
As in classical computer science, operations on qubits are described by gates. They describe well-defined actions that change the state of a qubit in a specific way. Physically, this is done, for example, by microwave pulses of a certain duration and field strength. The description by quantum gates is an abstraction from their concrete physical realization.

\begin{figure}[t]
	\centering
	\includegraphics[width=12truecm]{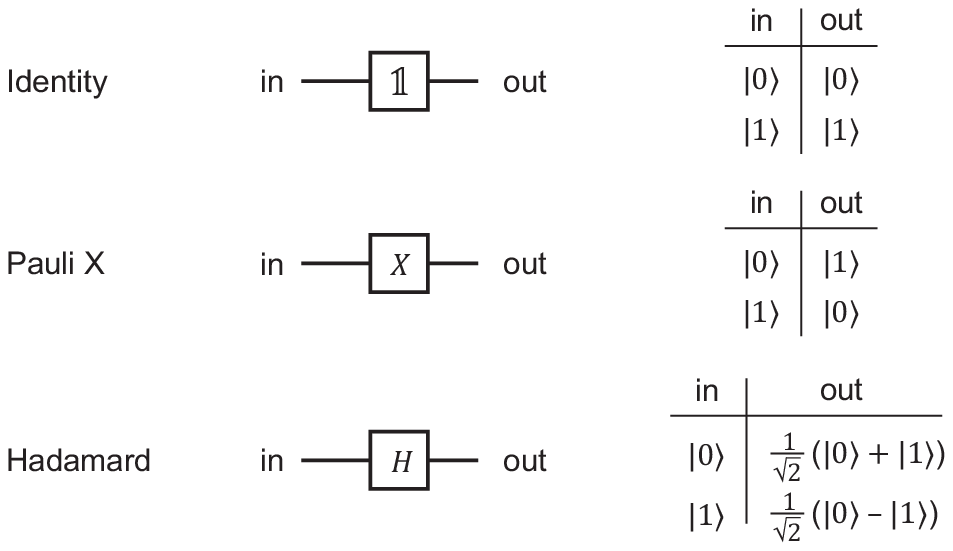}
	\caption{The quantum gates required for the Quantum Penny Flip algorithm}
	\label{abb5_8}
\end{figure}

For the realization of the Quantum Penny Flip algorithm, the gates shown in Fig.~\ref{abb5_8} are relevant: the identity operation $\mathbb{1}$ (which does not change the state), the quantum mechanical NOT gate (which, as already mentioned, is called the {Pauli $X$} gate) and the {Hadamard gate}. Further information on quantum gates can be found in books on quantum computing (e.\,g.~\cite{rieffelQuantumComputingGentle2011}) and numerous websites (e.\,g.~\cite{QiskitTextbook}).

\smallskip

{\bfseries Pauli $X$ gate: }
The Pauli-$X$ gate converts the basis state $\ket{0}$ to $\ket{1}$ and vice versa the basis state $\ket{1}$ to $\ket{0}$. It thus describes the flip of the coin. In Dirac notation it is given by the expression:
\beq X = \ket{0}\bra{1} + \ket{1}\bra{0}.\label{eq5_11}\eeq
We verify that the gate behaves according to the truth table in Fig.~\ref{abb5_8} by applying it, for example, to the basis state $\ket{0}$. We get:
\beq X \ket{0} = \Bigl[\ket{0}\bra{1} + \ket{1}\bra{0}\Bigr] \ket{0} .\eeq
Formally multiplying out the square brackets leads to the two scalar products $\braket{1|0}=0$ and $\braket{0|0} = 1$. Substituting this, we get as expected:
\beq X \ket{0} = \ket{1}.\eeq
Analogously, applying $X$ to the basis vector gives $\ket{1}$:
\beq X \ket{1} = \Bigl[\ket{0}\bra{1} + \ket{1}\bra{0}\Bigr] \ket{1} 
= \ket{0}\underbrace{\braket{1|1}}_{=1} + \ket{1}\underbrace{\braket{0|1}}_{=0} = \ket{0}.\eeq
Thus, the operator $X$ described by the expression~\eqref{eq5_11} acts on the basis vectors as shown in the truth table in Fig.~\ref{abb5_8}. 
Unlike the classical NOT gate, the Pauli $X$ gate is also defined for quantum mechanical superposition states. The resulting state follows directly from the linearity of the operation. If
$\ket{\psi_\text{in}} = \alpha \ket{0} + \beta \ket{1}$, then:
\beq  \ket{\psi_\text{out}} = X \ket{\psi_\text{in}} = \alpha \ket{1}  + \beta \ket{0}. \label{eq5_6}\eeq

\noindent
{\bfseries Hadamard gate:}
The Hadamard gate $H$ transforms the basis states $\ket{0}$ and $\ket{1}$ into superposition states:
\beq
H\ket{0} = \sqs \left(\ket{0} + \ket{1}\right) \quad \text{and} \quad H\ket{1} = \sqs \left(\ket{0} - \ket{1}\right).
\label{eq5_8}
\eeq
In quantum algorithms, it is one of the most commonly used gates because it generates superposition states. Superposition is the characteristic feature of quantum computers, and it is the prerequisite for quantum advantage. The Hadamard gate is therefore often used at the very beginning of a quantum algorithm.

In Dirac notation, the expression for the Hadamard gate is:
\beq H = \sqs \Bigl[ \ket{0}\bra{0} + \ket{1}\bra{0} + \ket{0}\bra{1} - \ket{1}\bra{1}\Bigr].\label{eq5_12}\eeq
The effect on the basis states can be checked in the same way as for the $X$ gate. With these two quantum gates, which are also important for more complex algorithms, we can discuss the winning strategy for Quantum Penny Flip.

\section{The winning strategy for Quantum Penny Flip}
We already mentioned that Alice is allowed to use all quantum gates in the quantum version of the game. We can now be more precise about that: she is allowed to use the Hadamard gate. Bob still has only the choice between flipping (corresponding to gate~$X$) or not flipping (corresponding to gate~$\mathbb{1}$).
Alice's winning strategy in Quantum Penny Flip can be formulated quite simply: Whenever it is her turn, she performs operation $H$ on the coin. If she follows this strategy, she wins -- no matter what Bob does.

\begin{figure}[b]
	\centering
	\includegraphics[width=12truecm]{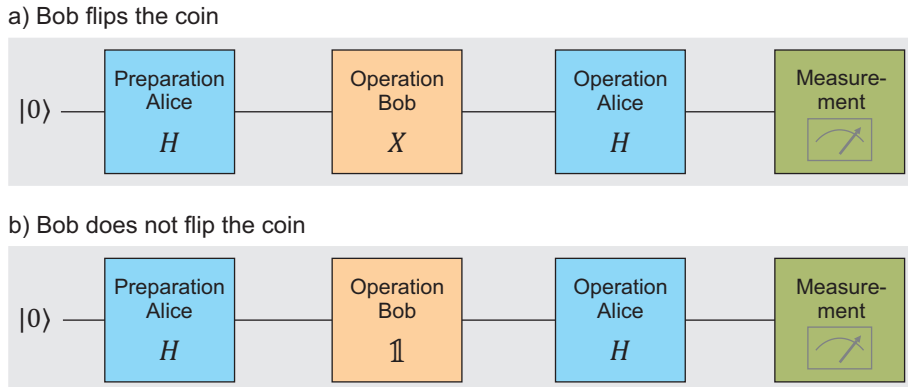}
	\caption{The sequence of possible qubit operations in the Quantum Penny Flip} 		
	\label{abb5_9}
\end{figure}

Fig.~\ref{abb5_9} shows the sequence of possible qubit operations. Initially, the qubit is prepared in the $\ket{0}$ state. As before, it is Alice's turn first, then Bob's, then Alice's again. If Alice follows her winning strategy and always uses $H$, there are two possibilities for the course of the game: Bob can flip the coin (operation $X$) or not (operation~$\mathbb{1}$).
Using the explicit expressions for the quantum gates given earlier, we can determine the outcome in either case.
\smallskip

\noindent {\em Case a) Bob flips the coin}\\
For the first case (Bob turns the coin), the final state of the qubit $\ket{\psi_\text{out}}$ is obtained by successively applying the operators $H$, $X$, and $H$ to the initial state $\ket{\psi_\text{in}} = \ket{0}$:
\begin{align} 
	\ket{\psi_\text{out}} = H X H \ket{0} = \sqs & \Bigl[ \ket{0}\bra{0} + \ket{1}\bra{0} + \ket{0}\bra{1} - \ket{1}\bra{1}\Bigr]\times \Bigl[ \ket{0}\bra{1} + \ket{1}\bra{0} \Bigr] \nonumber \\
	& \times \sqs \Bigl[ \ket{0}\bra{0} + \ket{1}\bra{0} + \ket{0}\bra{1} - \ket{1}\bra{1}\Bigr]\ket{0}\label{eq5_15a}
\end{align}
Multiplying out the terms in square brackets is made easier if we consider $\braket{0|1}=0$ and do not write down the corresponding terms at all. In this way we get:
\beq \ket{\psi_\text{out}} = \Bigl[ \ket{0}\bra{0} - \ket{1}\bra{1}\Bigr]\ket{0} = \ket{0}.\label{eq5_16a}\eeq
The result is $\ket{0}$ = heads, and Alice has won.
\smallskip 

\noindent {\em  Case b) Bob does not flip the coin}\\
The second possibility (Bob does not flip the coin) is even easier to handle. Since the unit operator $\mathbb{1} = \ket{0}\bra{0} + \ket{1}\bra{1}$ represents doing nothing, its presence does not affect the final result. Thus, two $H$ gates act on the qubit in succession.
\begin{align} 
	\ket{\psi_\text{out}} = H \mathbb{1} H \ket{0} = H  H \ket{0} = \sqs & \Bigl[ \ket{0}\bra{0} + \ket{1}\bra{0} + \ket{0}\bra{1} - \ket{1}\bra{1}\Bigr] \nonumber \\
	& \times \sqs \Bigl[ \ket{0}\bra{0} + \ket{1}\bra{0} + \ket{0}\bra{1} - \ket{1}\bra{1}\Bigr]\ket{0}\label{eq5_15}
\end{align}
The calculation follows the same scheme as above, and we get:
\beq \ket{\psi_\text{out}} = \Bigl[ \ket{0}\bra{0} + \ket{1}\bra{1}\Bigr]\ket{0} = \ket{0}.\label{eq5_16}\eeq
We see that applying the Hadamard gate twice returns to the initial state. The Hadamard gate is its own inverse. As a result, heads occurs also in this case, and again Alice has won.

In sum, we have found a definite winning strategy for Alice. No matter what Bob does -- flipping or not flipping the coin: If Alice applies the Hadamard gate on both moves, she cannot lose. This is an astonishing difference from the classical version of the game, where no winning strategy exists for either player. 

It is instructive to highlight the quantum mechanical background of this success: superposition states are a characteristic of quantum physics that is different from the classical case. The superposition state $\sqs \left(\ket{0} + \ket{1}\right) $ obtained by applying the Hadamard gate is invariant to the interchange of $\ket{0}$ and $\ket{1}$ -- i.e., to the application of the $X$ gate. 
Using this property of the superposition state is crucial to Alice's success and forms the basis of her winning strategy.

\section{Implementation on a real quantum computer}
\begin{figure}
	\centering
	\includegraphics[width=12truecm]{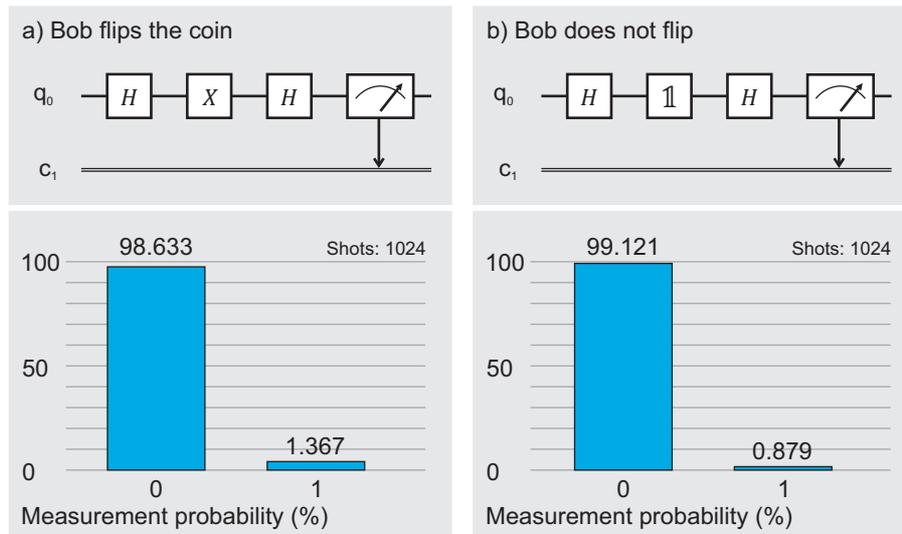}
	\caption{Result of the calculation on a quantum computer. At the end, the classical bit~c$_1$ contains the result of the quantum measurement at the qubit q$_0$.}
	\label{abb5_10}
\end{figure}

For the implementation of the Quantum Penny Flip algorithm we use the IBM's Quantum Composer~\cite{IBMQuantumComposer}.  
In this environment, programming is done graphically with symbols for the quantum gates (Fig.~\ref{abb5_10}, top). Individual qubits are represented by horizontal lines labeled $q_0, q_1, \dots$ (in the present case we need only one qubit $q_0$). The different gates can be placed on these ``wires'' by drag-and-drop. Each qubit is assumed to be prepared in the initial state $ \ket{0} $.

For learners, it may be a bit confusing at first that in the graphical representation the algorithm is read from left to right while the order of the operators in the bra-ket formalism is from right to left. For example, the operations $\ket{\psi_0}= X H \ket{0} $ are represented in the programming environment by $ q_0 ~\textendash \fbox{H} \textendash \fbox{X} \textendash $. 

The final measurement has to be explicitly specified in Quantum Composer.
It is symbolized by a stylized measuring device and can be placed in the same way as the gates. The additional classical bit represented by a horizontal double line (denoted $c_1$ in Fig.~\ref{abb5_10}) is used to store the measurement result. 

The symbolic representation of the gate arrangement for the two cases of our game is shown in Fig.~\ref{abb5_10}.
In Quantum Composer, the simulated state vector and the simulated measurement results are displayed immediately. However, the computation can also be performed on a real quantum computer. 
The results of such a computation for the two cases of our game are shown in Fig.~\ref{abb5_10}.
Due to decoherence, the results are affected by noise. Therefore, 1024 runs were performed in each calculation. The result confirms the winning strategy for Alice: except for the noise, in all cases the measurement at qubit q$_0$ yields the output 0 (= head). 

We also see from the results that each bit operation injects additional noise into the system. If Bob flips the coin, three operations have to be performed on the qubit. If he does not flip, it's only two operations (gate~$\mathbb{1}$ = do nothing). From the numerical values in Fig.~\ref{abb5_10} it can be seen that in the first case the deviation from the theoretically expected result is larger than in the second.

\section*{Acknowledgement}
This work has received funding by the European Union’s Horizon 2020 research and innovation programme under grant agreement No 951787.

\sloppy
\printbibliography
%
	

\end{document}